\documentclass[pre,twocolumn,superscriptaddress,amsmath,amssymb,showpacs,noshowkeys,a4paper]{revtex4}

\usepackage{amsmath}

\usepackage{graphicx}
\usepackage{dcolumn}
\usepackage{bm}
\usepackage[usenames]{color}
\usepackage{epstopdf}
\definecolor{mygray}{gray}{0.5}

\newcommand{\affmsc}{\address{Laboratoire Mati\`ere et Syst\`emes Complexes, Universit\'e Paris Diderot, CNRS - UMR 7057, B\^atiment Condorcet, 10 rue Alice Domon et L\'eonie Duquet, 75013 Paris, France, EU}}

\newcommand{\afflangevin}{\address{Institut Langevin, ESPCI ParisTech, CNRS - UMR 7587, 1 rue Jussieu, 75005 Paris Cedex 05, France, EU}}

\begin{document}
\title{Non-Hamiltonian features of a classical pilot-wave dynamics}

\author{M. Labousse}
\email{matthieu.labousse@espci.fr}
\afflangevin
\author{S. Perrard}

\email{stephane.perrard@univ-diderot-paris.fr}
\thanks{\\the authors contributed equally to this work}
\affmsc

\begin{abstract}
	A bouncing droplet on a vibrated bath can couple to the waves it generates, so that it becomes a propagative walker. Its propulsion at constant velocity means that a balance exists between the permanent input of energy provided by the vibration and the dissipation. Here we seek a simple theoretical description of the resulting non-Hamiltonian dynamics with a walker immersed in a harmonic potential well. We demonstrate that the interaction with the recently emitted waves can be modeled by a Rayleigh-type friction. The Rayleigh oscillator has well defined attractors. The convergence toward them and their stability is investigated through an energetic approach and a linear stability analysis. These theoretical results provide a description of the dynamics in excellent agreement with the experimental data. It is thus a basic framework for further investigations of wave-particle interactions when memory effects are included. 
\end{abstract}
\pacs{47 55.D- Drops,  05 45.-a, Nonlinear dynamics and chaos}
\maketitle

\section{Introduction \label{Sec:I}}
	From the simple pendulum to more complex oscillators, the unavoidable coupling with the environment generates dissipation. Even if the loss of energy is balanced by a permanent input, the Hamiltonian structure is lost. The description of open systems, i.e. when the coupling with the environment cannot be neglected, has been explored in the framework of self-sustained oscillators. A theoretical approach was first developed by van der Pol \cite{VanDerPol_1926} to describe the spontaneous regular heartbeats \cite{VanDerPol_1928}. This oscillator is characterized by an amplitude-dependent friction term: the oscillation is amplified for a small amplitude and damped for a larger one. This equation has been widely studied as a rare example of nonlinear equations that can be solved analytically. \\
	
	Self-propelled entities also result from a coupling with their environment, from the motion of living animals to artificial self-propelled systems. The efficiency of the propulsive mechanism depends on the permanent input of energy and the specificity of the dissipative process. The origin of the propulsion is diverse \cite{Schmidt-Nielsen_1972}, from low Reynolds propulsion of a flagellum \cite{Purcell_1997} to the bird flight, from vertically vibrating anisotropic disks \cite{Deseigne_2010} to motile colloids \cite{Bricard_Nature_2013}. However, all these situations can be described by a speed-dependent friction term, as initially introduced by Lord Rayleigh \cite{Rayleigh_1877}. \\

\begin{figure}[t]
\begin{centering}
\includegraphics[width=0.8 \columnwidth]{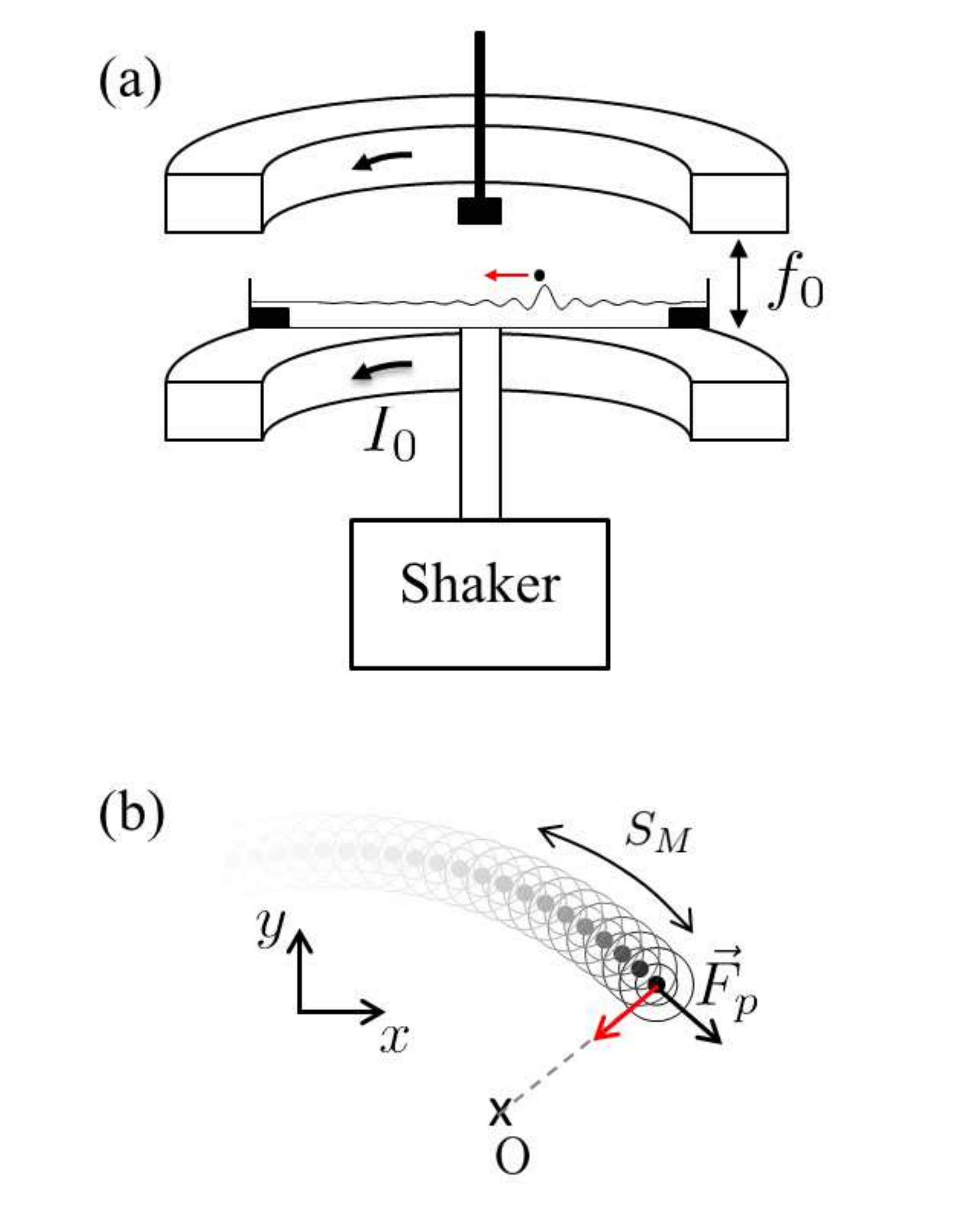}
\caption{(Color online) Sketch of the experimental set up. (a) A bath of silicon oil is set on a vertically oscillatory motion at a frequency $f_0=80$ Hz. A drop deposited on the surface starts bouncing without coalescence (prevented by the air layer between the liquid surface and the drop). A small amount of ferrofluid is then encapsulated inside the drop so that it becomes sensitive to a magnetic field. A current $I_0$ through two coils in a Helmholtz position generates a homogeneous magnetic field at the bath surface. By adding an additional magnet at a distance $d$ above the surface, an attractive force can be generated on the drop. The resulting potential can be modeled by a harmonic well: $E_p= - \kappa r^2/2$ (red arrow online). (b) Details of the walker's motion. The walker moves at a velocity $V_0$ due to the propulsion by damped Faraday waves. These waves (indicated by large open circles in the figure) are generated by the impact of the drop (indicated by small, solid circles) and are damped with a characteristic time $\tau$ larger than the vertical period of motion. Their decay time $\tau$ can be converted in a memory length $L=\tau V_0$. Here, we study the limit in which $L$ is small compared to any curvature radius of the trajectory.\label{fig1}}
\end{centering}
\end{figure}
	A simple oil drop can be propelled as well. Set on a vertically oscillating liquid bath, a millimetric drop does not necessarily coalesce due to the presence of an air film between it and the liquid surface \cite{Walker_nature}. The coupling with the environment arises from the Faraday waves. They appear spontaneously at the surface of a vibrated liquid bath by a parametric instability mechanism \cite{Benjamin_1954}, for an acceleration amplitude larger than the Faraday threshold $\gamma_F$. Just below this threshold, the drop impacts excite slowly damped Faraday waves \cite{Eddi_JFM_2011}. As a result, the drop bounces on a surface perturbed by the previously generated waves. A horizontal momentum can be transmitted to the bouncing drop which propels it \cite{Walker_nature,Dorbolo_NJP_2008,Terwagne_PM_2008}. The resulting entity, called a walker, is thus formed by a localized drop and an extended accompanying wave packet. This association is a rare example of a macroscopic dual object formed by a particle and its associated wave. Walkers display various quantum like effects: diffraction through a slit \cite{Couder_diffraction}, tunneling effect \cite{Eddi_Tunnel},  Zeeman-like splitting \cite{Eddi_PRL_2012}, Landau level analog \cite{Fort_PNAS_2010}, and surprising behaviors in cavity \cite{Shirokoff_Chaos_2013,Harris_PRE_2013}. Recently, the confinement of walkers in a two-dimensional harmonic potential leads to a set of discrete eigenstates for the possible trajectories and their associated wave field \cite{Perrard_natureC_2013}. In spite of the similarity between walkers and some quantum systems \cite{Couder_JP_2011,Bush_PNAS_2010}, the dynamics of a walker is far from the quantum world in many aspects. By its periodic interaction with the liquid surface, a walker is an open system, in which the interaction with the environment plays a crucial role. In particular, the underlying hydrodynamic equations governing the bouncing states \cite{JFM_Suzie,Gilet_PRE_2007,Molacek_JFM_1_2013,Molacek_JFM_2_2013} are not conservative, and the generation of the capillary waves dissipates energy. This energy loss is compensated by the propulsion from the standing Faraday waves at each impact. Therefore the walking state is the result of an energy balance between propulsion and dissipation as any self-propelled entity. \\

	In this paper, we identify the consequences of the self-propulsion mechanism on the horizontal walker dynamics. We investigate its horizontal dynamics with a model description of the walker propulsion based on a Rayleigh-type friction law. We focus on a walker trapped by a harmonic potential well. The study is also restricted to the short memory regime, i.e, when the damping time $\tau$ of the Faraday waves is much shorter than the period of rotation of the walker in the harmonic well.  Section \ref{Sec:II} describes the experimental results in the harmonic well, in the limit of short memory. Section \ref{Sec:IIA} is devoted to the experimental set up. Section \ref{Sec:IIB} presents the general features of the walker motion in this harmonic well. In Sec. \ref{Sec:IIC}, the transients' behaviors are investigated. Based on this description, we construct in Sec. \ref{Sec:III} a general structure for the equation governing the propulsion of the walker at low memory. The simplified equations of motion are derived in Sec. \ref{Sec:III}A. Then, the theoretical model is validated with the experimental data in Sec. \ref{Sec:IIIB}. Section \ref{Sec:IIIC} deals with an energetic approach for analyzing the convergence towards the attractors of the dynamics. Finally a linear stability analysis close to these attractor points is presented in Sec. \ref{Sec:IIID}. As a conclusion, we highlight some general consequences of the self-propulsion process for the walker dynamics.
\section{Walker trapped in a harmonic well: experiments \label{Sec:II}}
\subsection{Experimental set up \label{Sec:IIA}}
	 As sketched in Fig. \ref{fig1}(a), a bath of silicon oil of viscosity $20$ cP is set into an oscillatory motion vertically at a frequency of $80$~Hz and at an acceleration amplitude $\gamma_m$. A stationary wave pattern called Faraday waves appear spontaneously at the surface above an acceleration threshold $\gamma_F = 4.5 \pm 0.1g$ where $g$ denotes the acceleration of gravity. The appearance of Faraday waves is triggered by a parametric instability at half the forcing frequency so the wave pattern oscillates vertically at a period $T_F=0.025$ s \cite{Benjamin_1954}. The usual dispersion relation of surface waves prescribes the corresponding wavelength $\lambda_F= 4.75$ mm.
	 
	A submillimetric drop is placed on the vibrating bath tuned slightly below the Faraday threshold. At a typical bath acceleration of $4g$, the drop bounces at twice the period of the bath, which corresponds to the Faraday period $T_F$, so that the drop becomes a Faraday wave exciter. At each impact, the drop generates a capillary front which leads to the formation of a circular standing Faraday wave pattern centered at the impact point. The complete mechanism of the wave generation is detailed in \cite{Eddi_JFM_2011,Molacek_JFM_2_2013}. The amplitude of the emitted wave decreases in time as the bath acceleration amplitude $\gamma_m$ is chosen slightly below the instability threshold. The dimensionless parameter $M=\tau/T_F$ is a measure of the persistence time. It compares the damping time $\tau$ of the Faraday waves to the time $T_F$ between two successive drop bounces. For a memory parameter larger than 1, the Faraday wave emitted by each bounce is still surviving during at least the next bounce. The liquid surface will be thus perturbed by the waves generated during the $M$ previous impacts. The memory parameter $M$ can be tuned through the difference between the bath acceleration $\gamma_m$ and the Faraday threshold $\gamma_F$. Indeed, close to the Faraday threshold,  the damping time diverges as $M \propto \gamma_m/(\gamma_F-\gamma_m)^{-1}$ \cite{Eddi_JFM_2011,Molacek_JFM_2_2013}.
	
	The propulsion mechanism is a direct consequence of this memory effect. As soon as the decay time of Faraday waves is larger than the Faraday period ($M>1$), the drop hits a surface perturbed by the previous impacts. Since the drop bounces on an inclined surface, there is a transfer of horizontal momentum to the drop. It has been shown that this situation leads to a pitchfork bifurcation at $M\approx 3$ \cite{Walker_nature,Oza_JFM_1_2013}. The vertical bouncing state becomes unstable and the drop starts moving, propelled by the waves emitted in its near past. 

	In the present study, the acceleration amplitude is set at about 90\% of the Faraday threshold $\gamma_F$, corresponding to a decay time $M \approx 7$. For such an acceleration, bouncing drops of diameter $D = 600 \pm 50$ $\mu$ m moves at a horizontal velocity $V_0$ averaged on the vertical period of motion between 5 and 15 mm/s. The exact velocity value depends on the drop size and the bath acceleration and is fixed for a given drop \cite{JFM_Suzie}.
			
	For each drop, the decay time can be converted to a memory length $S_M=M V_0 T_F$. This length $S_M$, sketched in Fig. \ref{fig1}(b), corresponds to the typical distance along which the Faraday wave sources are still active behind the drop. In this article, we consider the low memory regime, i.e., when the memory length $S_M$ is smaller than the typical curvature radius of any curved trajectory followed by the walker. One important consequence is that the wave sources contributing to the propulsion can be considered as mainly aligned on a straight line behind the drop. The force exerted by each wave on the drop is thus mainly tangential to the trajectory. Consequently the global resulting force is mainly tangential to the trajectory.
		
	In the absence of other external force, a drop bounces along a straight line until it reaches the boundary of the container. A more interesting experimental case is obtained when the walker is confined. This situation is obtained by trapping the walker in a harmonic potential well as sketched in Fig. \ref{fig1}(a). For this purpose, the millimetric drop is loaded with a small amount of ferrofluid and is exposed to a magnetic field. Two coils in a Helmholtz configuration provide a uniform magnetic field in the region of interest and induces a magnetic moment $m_B$ to the drop. An additional non uniform magnetic field is generated by a cylindrical magnet with a diameter of $15$ mm and a height of $5$ mm placed above the liquid bath. The interaction between the external magnetic field and the drop magnetic dipole generates an external force on the magnetized drop. When the vertical distance $d$ between the magnet and the liquid surface is larger than the magnet diameter, the potential well can be considered as harmonic \cite{Perrard_natureC_2013}. For a drop of mass $m$, the dynamics is then reduced to the motion of a bouncing particle propelled by a wave with an external forcing $\mathbf{F}_{\mathrm{ext}}=-\kappa \mathbf{r}$. $\kappa$ is a spring constant which can be tuned by changing the distance $d$. The vector $\mathbf{r}$ denotes the distance from the drop center to the symmetry axis of the magnet. Additional experimental details have been given in the Supplemental Material of a previous work \cite{Perrard_natureC_2013}.
	
\subsection{Circular motion \label{Sec:IIB}}
	\begin{figure}[t]
\begin{centering}
\includegraphics[width=1 \columnwidth]{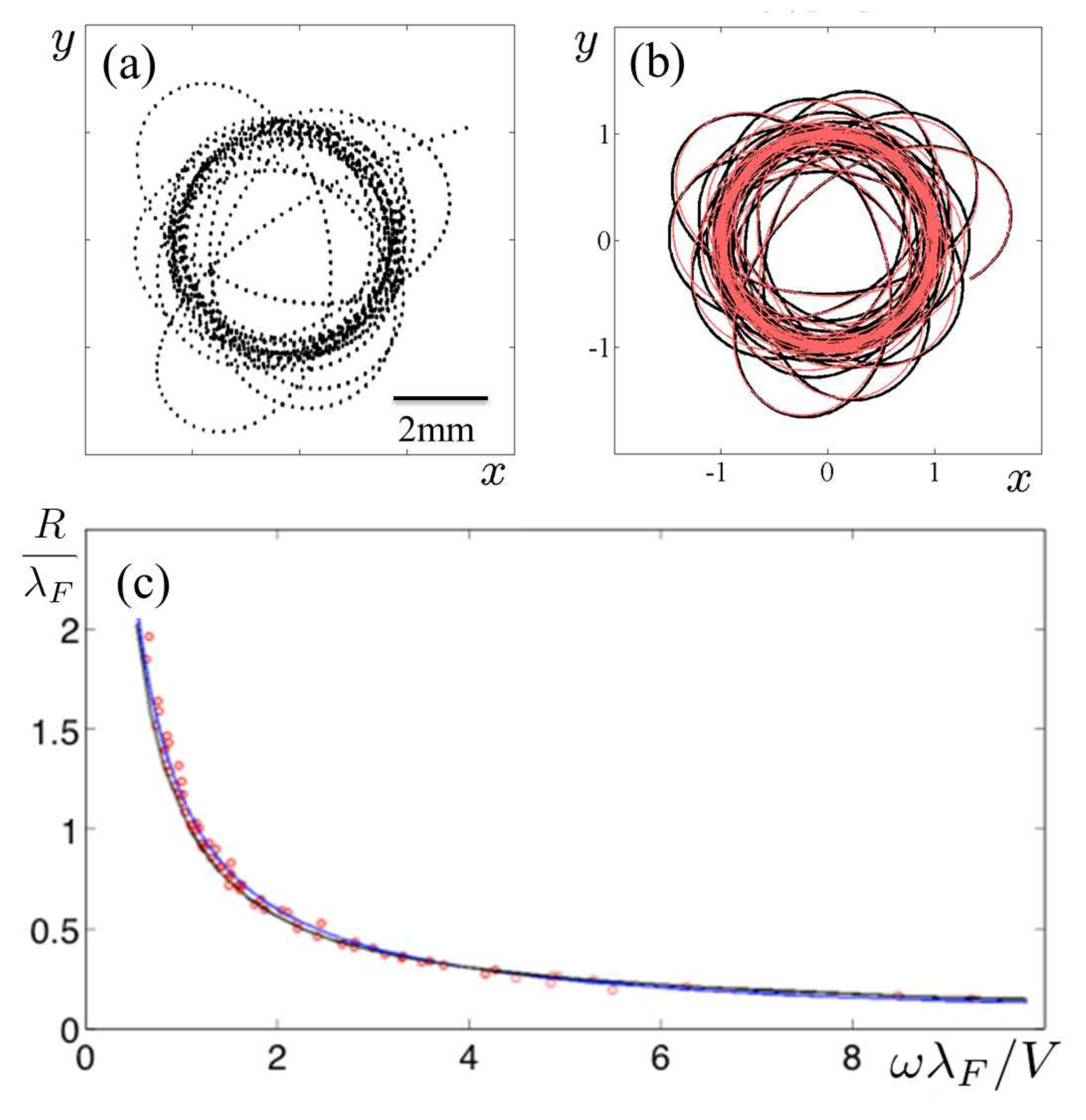}
\caption{(Color Online)(a) Chronophotography of the transient motion of the walker being trapped by the magnetic field. The time increment is fixed at 0.05 s which corresponds to one point every two bounces. After a few oscillations, the walker trajectory converges to a circular orbit. (b) Transient regimes obtained with Fort's model of the walker dynamics (solid black line) and numerical solving of the Rayleigh equation with $\Gamma=25.5$ [solid red line (online), light gray (printed)]. The quantitative agreement between experiments and numerical simulations shows that the walker propulsion can be described through a friction term depending on the velocity. (c) Orbit radius $R/\lambda_F$ as a function of the dimensionless frequency $\lambda_F \omega / V$, where $\omega$ is the characteristic frequency of the drop in the harmonic well. Experiments for two drops of velocity $V$=10mm/s and 8 mm/s  (\color{red}{$\circ$}).\color{black} The experimental data (red dots online) are compared to the simple scaling $R/\lambda_F=V/\omega\lambda_F$ (solid black line downer). The agreement is good, without use of any fit parameter. The blue (upper) line (blue online, gray printed) indicates the result of Fort's numerical model. \label{fig2}}
\end{centering}
\end{figure}
	When the walker is released from the edge of the cell, the magnetic force pulls it towards the center of the harmonic potential. As the drop is self-propelled, it will never stop at the center but orbits around it. This trapping mechanism is represented on a chronophotograph as seen from above [see Fig. \ref{fig2}(a)]. The position is sampled every $0.05$ s, corresponding to the time interval of two vertical periods $T_F$. In a few oscillations, the trajectory converges to a circular motion in which the walker orbits around the center at its free velocity $V_0$. Such a trajectory can also be computed from the numerical model developed by E. Fort \cite{Eddi_JFM_2011,Fort_PNAS_2010} to mimic the walker dynamics; [see Fig. \ref{fig2}(b)]. The same quantitative behavior is recovered with this numerical model in Fig. \ref{fig2}(b) (solid black line). Finally, balancing the centripetal acceleration  $-m V_0^2/R$ with the spring force $-\kappa R$, scales the final orbit radius $R/\lambda_F$ as $\Lambda=(\lambda_F \omega/V_0)^{-1}$ where $\omega=\sqrt{\kappa/m}$ corresponds to the natural frequency of the magnetic potential well, (see figure \ref{fig2}c). The coefficient of proportionality $a=(R/\lambda_F)/\Lambda\simeq 1.2 \pm 0.05$ differs slightly from $1$ as investigated by Oza \textit{et al.} \cite{Oza_JFM_2_2013} and interpreted as a contribution of the surface waves in the radial mechanical balance.
	
	For a Hamiltonian system, one would expect a more diverse set of trajectories in a two dimensional harmonic well. But among all possible solutions, only the circular motion enables the particle to move at a constant speed. The selection of particular trajectories is a consequence of the dissipative nature of the walker dynamics. 

\subsection{Convergence to circular attractors \label{Sec:IIC}}	
\begin{figure}[t]
\begin{centering}
\includegraphics[width=0.9 \columnwidth]{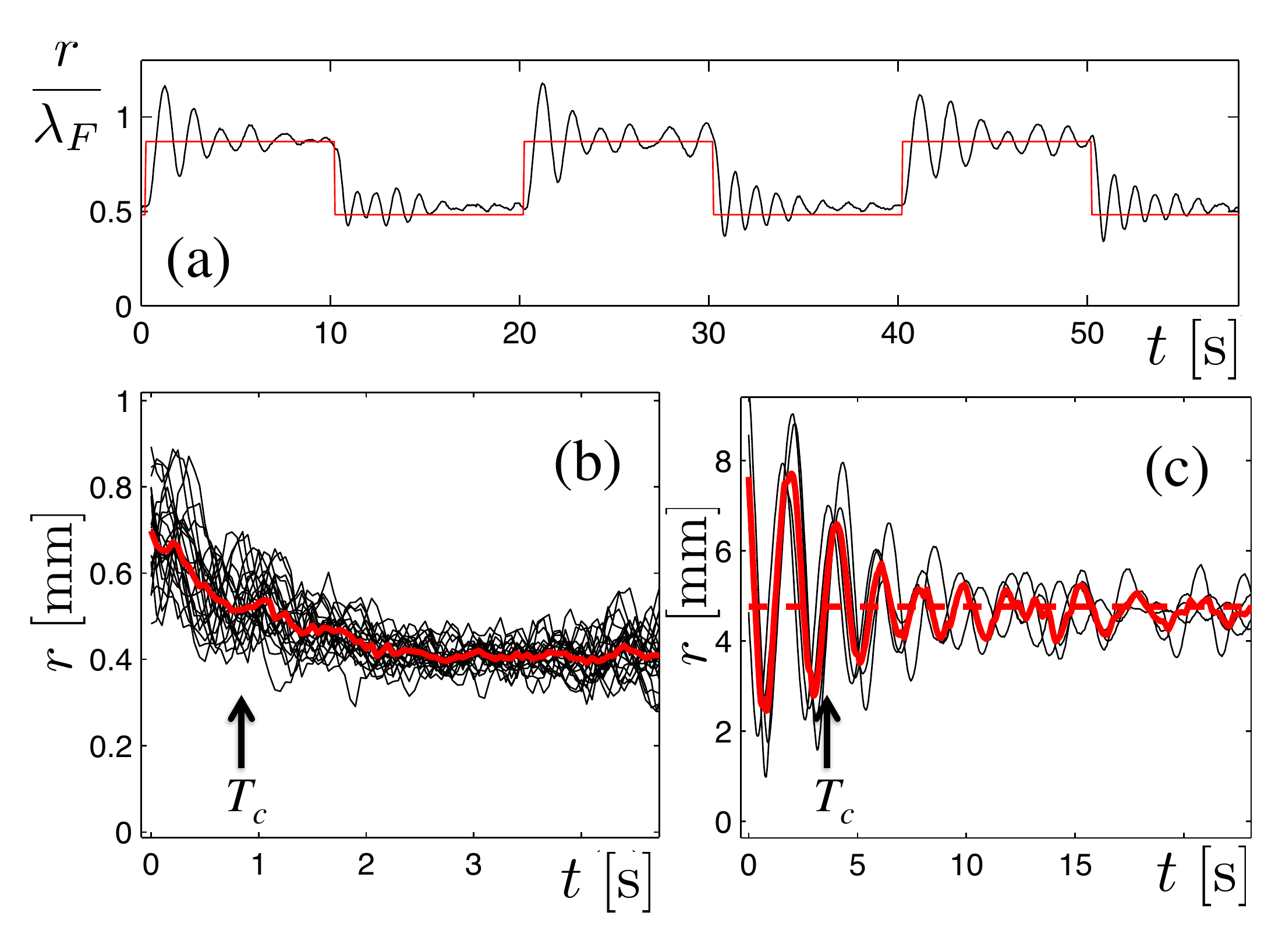}
\caption{(Color onlie) Experimental characterization of the transient behavior. The magnetic force changes abruptly every ten seconds, giving rise to repeated transients to the asymptotic radius. For each time interval, the experimental radius (oscillating black solid line) and the expected asymptotic radius (solid red line online) are shown. (b) Superposition of ten experimental transients for a final imposed radius of $R_1=0.4$ mm (black thin solid lines) corresponding to an imposed $\omega=\omega_1$. The red thick solid line indicates the average of the experimental transients. The radius converges by oscillating to the attractor without overshoot. (c) Superposition of ten experimental transients in the case of a final imposed radius of $R_2=4.8$ mm (thin black solid lines) corresponding to an imposed $\omega=\omega_2$. The red thick solid line indicates the average of the experimental transients. The radius converges to the attractor through damped oscillations.\label{fig3}}
\end{centering}
\end{figure}
	In dissipative systems, any volume of the phase space can contract in time and additional measurements have been carried out to study the transient regime. One way to proceed, is to study the transition between circular motions of different radius by abruptly changing the external force. A walker is then prepared on a circular orbit and the current inside the coils is switched between two values $I_1$ and $I_2$. The consequence is an abrupt change of the characteristic frequency value from $\omega_1$ to $\omega_2$. A typical experimental signal is shown in Fig. \ref{fig3}(a). We observe a transition between two different circular motions of radii $R_1$ and $R_2$ corresponding to the two values of the spring constant. After each switch, the walker follows a reproducible transient regime and converges to the new circular attractor. This transition is repeated ten times with a duration of 10 s between each switch, larger than the observed transient duration. For small orbits, the radius decreases slowly to the final state [see Fig. \ref{fig3}(b)]. For larger orbits, the radius increases quickly, crosses the final value and follows damped oscillations around the final state [see Fig. \ref{fig3}(c)]. The transient behavior depends only on the final state which can be reached either from a smaller or a larger orbit. We thus find for the evolution of the instantaneous radius $R(t)$, typical transient behavior of a second-order oscillator, in which the quality factor strongly depends on the orbiting period $T_R=2 \pi/\omega$ of the final orbit. The qualitative change of the transient behavior shows that at least two time scales control the transition. Depending on their ratio, the transient will evolve from  oscillations without overshoot (small orbits) to damped oscillations with overshoot (large orbits). The first time scale $1/\omega$ is imposed by the external force, and should be compared to another one. Where does this second time scale come from? We address this question in the next paragraph through a theoretical approach.
			
\section{A simple model of self-propelled drops: the Rayleigh equation \label{Sec:III}}
\subsection{Self-propulsion model \label{Sec:IIIA}}
    The second time scale relies on the dissipative nature of the motion. In the present case, the energy exchange comes from the coupling of the drop to the bath in two ways. First, the drop loses energy at impact by viscous dissipation but also because a new propagating capillary wave is generated. Second, when the drop hits a non horizontal part of a wave, its horizontal momentum is changed. Once the walker moves at its free velocity, i.e., in a permanent regime, these two terms balance one another. A phenomenological description of the propulsive force has been introduced by A. Boudaoud \cite{Walker_nature}, E. Fort \textit{et al.} \cite{Fort_PNAS_2010}, and extended by Oza \textit{et al.} \cite{Oza_JFM_1_2013}. Experiments show that the fluctuations of speed are typically smaller than $20$\%. This speed constraint enables us to introduce a simplified expression to model the net accelerating force $\mathbf{F}_p$. We derive it through symmetry arguments and show \textit{a posteriori} that such an expression fits the experimental results. As we are in the short memory regime, the force $\mathbf{F}_p$ is mainly tangential to the trajectory, i.e., $\mathbf{F}_p=  F_p\mathbf{V}/\Vert\mathbf{V} \Vert$. $F_p$ must only depend on the amplitude of the speed $V$ and not on its direction. An exchange $V\rightarrow -V$ should only change the sign of $F_p$. Consequently $F_p$  must be an odd function in $V$, vanishing for a velocity equal to the equilibrium one $V_0$. The simplest expression satisfying these requirements is 
\begin{equation}
\mathbf{F}_p=-\gamma_0 \left(\left(\frac{V}{V_0}\right)^2-1\right) \mathbf{V} +O(V^5),
\label{Forcepropulsion}
\end{equation}
with $\gamma_0$ a constant homogeneous to a friction coefficient. For $V<V_0$ the force is propulsive whereas it acts as a friction term for $V>V_0$. This force vanishes for $V=V_0$. Such a term is called a Rayleigh-type friction \cite{Rayleigh_1877} and is commonly used to model the motion of active particles. It is the lowest order expansion of a force which depends on the velocity and vanishes for an equilibrium velocity. 

	Alternatively, Eq. \ref{Forcepropulsion} can be derived by a direct calculation detailed in the Appendix. This form may also be derived from the integrodifferential formulation developed by Oza \textit{et al.} \cite{Oza_JFM_1_2013}. We state that $\mathbf{F}_p$ is mainly tangential, but the corrective radial term could also be added. Bush \textit{et al.} \cite{Bush_JFM_2014} shows with a higher order expansion from an integrodifferential form that this extra term is responsible for the prefactor $a$ discussed in Sec. \ref{Sec:IIB}. Note that the various expressions of this velocity-dependent term introduced in previous studies \cite{JFM_Suzie,Molacek_JFM_2_2013} are all equivalent to the Rayleigh friction provided that the fluctuations in speed remain small, i.e., less than 20\% of the free velocity $V_0$. The 
Rayleigh-type friction retains the minimal complexity required to describe the walker propulsion theoretically. However, our expression is not intended to describe the consequence of an abrupt change of velocity. Such cases have been encountered for very specific situations, like highly disordered trajectories \cite{Harris_PRE_2013}, where velocity fluctuations become larger than $50$\%. In the present article, we do not observe significant discrepancies between our simplified model and the experimental data.
	
	The horizontal dynamics can then be rationalized by a continuous equation as the bouncing period $T_F=1/40s$ is much smaller than the period of rotation $T_R>1$ s \cite{Oza_JFM_1_2013}. Using the Rayleigh-type friction law and the continuous limit, the simplified equation of motion for a walker in a harmonic potential becomes:
\begin{equation}
\frac{d \mathbf{V}}{dt}=-\omega^2 \mathbf{r} -\frac{1}{T_V} \mathbf{V}\left(\left(\frac{\mathbf{V}}{V_0}\right)^2 - 1\right)
\label{equadiff1}
\end{equation}
with $T_V=m/\gamma_0$ a characteristic time. The one-dimensional case is equivalent to a van Der Pol oscillator for the velocity, signature of the Hamiltonian structure breaking. In spite of its apparent simplicity, only a few articles deal with the two-dimensional case. Erdmann \textit{et al.} \cite{Erdmann_2005} studied this equation in two dimensions including noises, indicating that circular motions are solutions. However, the two dimensional Rayleigh equation has not been solved analytically. This can be attributed to the friction term, which involves the norm of the speed. This term provides a coupling between the motions in the two directions of space. 

\subsection{Validation of the theory \label{Sec:IIIB}}
One particularity of this equation is the presence of two different time scales. The external force provides the period of oscillation while the Rayleigh type friction introduces another time scale $T_V$, the typical time needed for a walker to recover its free velocity. It should not depend on the external force and is thus a physical parameter for a given bouncing drop at a given memory $M$. Even if symmetry arguments enable us to obtain the simplified equation \ref{equadiff1}, they do not provide the value of this characteristic time $T_V$. It will be determined by comparing the model with the experimental data. \\

At this step, we introduce the dimensionless position $\mathbf{r} \rightarrow \mathbf{r} \omega/V_0=(x,y)$ and speed $\mathbf{v}=\mathbf{V}/V_0$ while the time is scaled by the natural frequency $t \rightarrow \omega t$.  The equation of motion in Cartesian coordinates yields
\begin{equation}
\left\{
    \begin{array}{ll} 
  \displaystyle     \ddot{x}=- x+\Gamma\dot{x}\left(1-\left(\dot{x}^2+\dot{y}^2\right) \right)\\
  \displaystyle     \ddot{y}=- y+\Gamma\dot{y}\left(1-\left(\dot{x}^2+\dot{y}^2\right) \right) 
     \end{array}
\right.
\label{equationdyninertia}
\end{equation}
here the upper dots indicate a derivative with respect to the dimensionless time. $\Gamma=1/(\omega T_V)$ is the natural dimensionless parameter comparing the two relevant time scales. \\

\begin{figure}[t]
\begin{centering}
\includegraphics[width=0.7 \columnwidth]{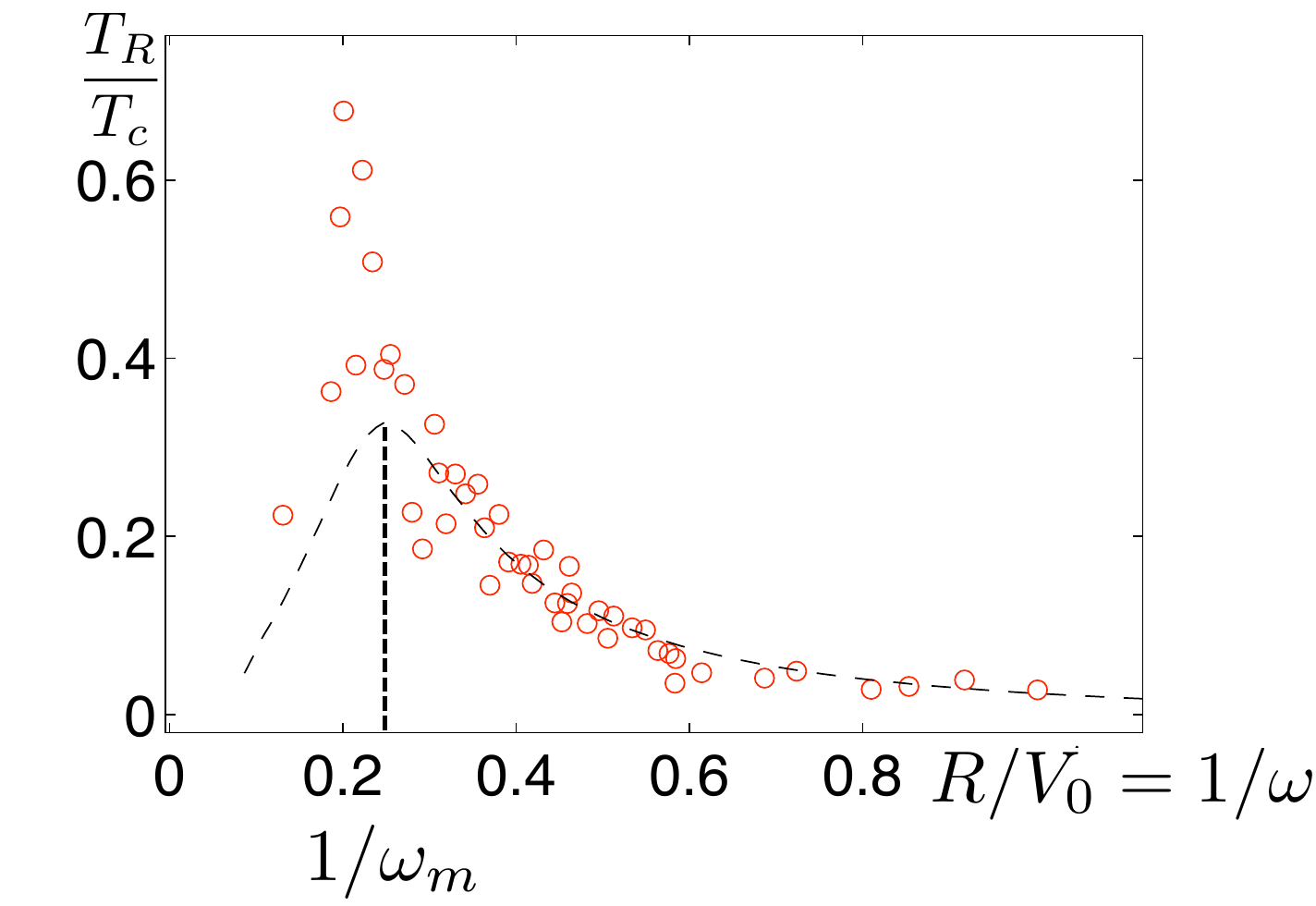}
\caption{(Color online) Dimensionless characteristic rate $T_R/T_c$ of convergence to the circular motion ($T_R$ is the orbiting period) as a function of  $R/V_0$: experimental data (red circle online), numerical solving of the Rayleigh oscillator (black dashed line). A larger value means a faster convergence.  Numerically, we change $\omega=1/(\Gamma T_V)$ by decreasing $\Gamma$ with $T_V=0.14 \pm 0.08$ s. A minimum in the convergence time is observed both in the experiment and in the theory.\label{fig4}}
\end{centering}
\end{figure}     
	As the full time-dependent resolution is not of particular importance for the current study, a numerical implementation of this system of equations has been performed. The solution of the Rayleigh equation can be compared to both experiments and simulations from Fort's model. The transients for $\Gamma=25.5$ are compared in Fig. \ref{fig2}(b) and shows a good agreement. The theoretical final radius can also be compared with the experimental and numerical results [see Fig. \ref{fig2}(c)] and are also in good agreement. \\   	

	The convergence to the final radius, oscillating or not [see Figs \ref{fig3}(c) and \ref{fig3}(b)] can be approximated by an exponential law and we define a converging time $T_c$ corresponding to this exponential decay time. Thus, this converging time can be evaluated directly from the time evolution of the radius both for experimental and numerical data. Figure \ref{fig4} indicates by red circles, the experimental evolution of the inverse of the converging time $T_c$ with the orbital frequency $\omega=V_0/R$. We also indicate by dashed black lines, the prediction of Eq. \ref{equationdyninertia} as we increase $\omega=1/(\Gamma T_V)$, i.e., by decreasing $\Gamma$. $T_V$ can be seen as a free parameter obtained from the best matching between the experimental data and the theory. It enables us to compare the experimental data of convergence with the theoretical predictions and gives $T_V=0.14\pm 0.01$ s. The existence of an optimal value of convergence for $1/\omega_m=0.25$ s.rad$^{-1}$ or equivalently $\Gamma_m=1.75 $ can be seen in Fig. \ref{fig4}. For $\omega \gg \omega_m$ (or $\Gamma \ll \Gamma_m$), the nonlinear terms become small and the dissipation averaged on one period is limited: the time of convergence will naturally increase. In the opposite case $\omega \ll \omega_m$ (or $\Gamma \gg \Gamma_m$), the two characteristic time scales differs strongly so that the nonlinear term is not efficient to reach the equilibrium speed at $v=1$. In both cases, the consequence is an increase of the convergence time. We note that the convergence time depends slightly on the initial conditions as soon as the initial motion is not radial. In this purely radial case, the drop starts oscillating along a straight line passing close to the center. But the nonlinear coupling $\dot{x}^2+\dot{y}^2$ makes this regime unstable to transverse perturbations and the dynamics converges finally to a circular motion.\\
	
 	We have investigated the behavior of the walker in a short memory limit and have shown that the dynamics is well approximated by a two-dimensional Rayleigh oscillator. We have demonstrated that there are two regimes of convergence depending on the relative magnitude of the two time scales of the dynamics. By comparing experimental results and model calculations, we have been able to obtain the value of the second characteristic time $T_V$. We now turn to the question: which role does each mechanism play to stabilize onto the circular attractors?

\subsection{Nonlinear stability \label{Sec:IIIC}}
\begin{figure}[t]
\begin{centering}
\includegraphics[width=0.8 \columnwidth]{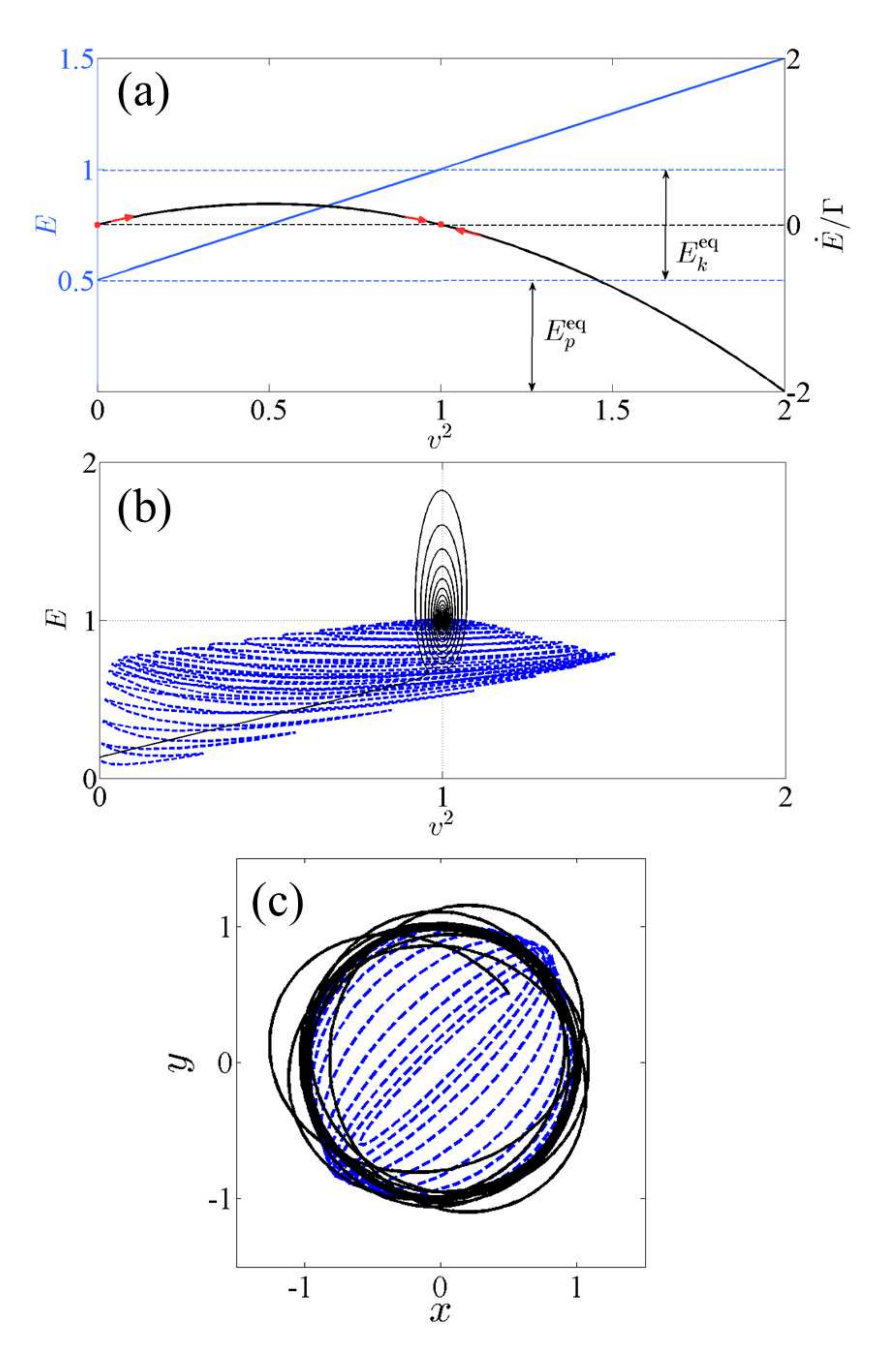}
\caption{(Color online) Stability of the circular attractor. (a) Schematics to illustrate the stability of the limit cycle $(r=r_0=1,v=v_0=\pm 1)$. The evolution of $\dot{E}$ with the square of the dimensionless speed $v^2$ (in black, right $y$-axis). The evolution of the energy $E$ with $v^2$ for a given dimensionless radius $r=1$ (in blue online, light gray printed, left y-axis). Two fixed points $\dot{E}=0$ are identified: $v=0$ unstable [$d\dot{E}/d(v^2) >0$] and $v=\pm 1$ stable [$d\dot{E}/d(v^2) <0$]. At the equilibrium $(r=1,v=1)$, there is an equipartition of energy between the kinetic energy $E_k^{\mathrm{eq}}$ and the potential energy $E_p^{\mathrm{eq}}$. (b) Time evolution of the total energy as a function of $v^2$. At large time, the dynamics reaches the fixed point $(E=1,v=\pm1)$, i.e., $(r=1,v=\pm 1)$. The case of large $\Gamma$ is in the solid back line $(\Gamma=70$) while the blue dashed line indicates the case of small $\Gamma=0.1$. (c) Trajectories corresponding to the case in part (b).\label{fig5}}
\end{centering}
\end{figure}
	 As shown in the previous part, the transient regime of a walker can present either oscillations of radial position or oscillations of speed. The time of convergence is minimal when their relative amplitude of oscillations is comparable. $\Gamma_m$ can be seen as the value of $\Gamma$ which optimizes the exchange between potential and kinetic energy. For all other values, the mismatch of the two timescales increases the time of convergence. \\
	The oscillations of both position and velocity with variable amplitudes suggest studying the stability through an energy approach. For this purpose we denote the dimensionless mechanical energy $E=v^2/2+r^2/2$. It evolves in time since the system is not conservative. Equations \ref{equationdyninertia} yields
\begin{equation}
\dot{E}=\Gamma v^2\left(1-v^2 \right).
\end{equation}
The energy evolves as a function of the speed only, and its three fixed points are $v=0$ and $v=\pm 1$. Only the solution $v=v_0=\pm 1$ and the related condition $\dot{E}=d(v_0^2/2+r^2/2)/dt=dr^2/dt=0$ defines an attractive limit cycle as sketched in Fig. \ref{fig5}a. This equilibrium corresponds to an equipartition of the total energy into kinetic $E_k^{\mathrm{eq}}$ and potential $E_p^{\mathrm{eq}}$ terms. The equipartition of the energy is not necessarily a property of a non-Hamiltonian system. In the plane $(E,v^2)$, two different transient regimes are represented in figure \ref{fig5}b for two different values of $\Gamma$ and the related paths in Fig. \ref{fig5}c. The case $\Gamma/\Gamma_m \gg 1$ (solid black line) presents limited oscillations in speed while the case $\Gamma/\Gamma_m \ll 1$ corresponds to oscillations in speed and position (dashed blue line).

\subsection{Linear response \label{Sec:IIID}}
\begin{figure}[t]
\begin{centering}
\includegraphics[width=1 \columnwidth]{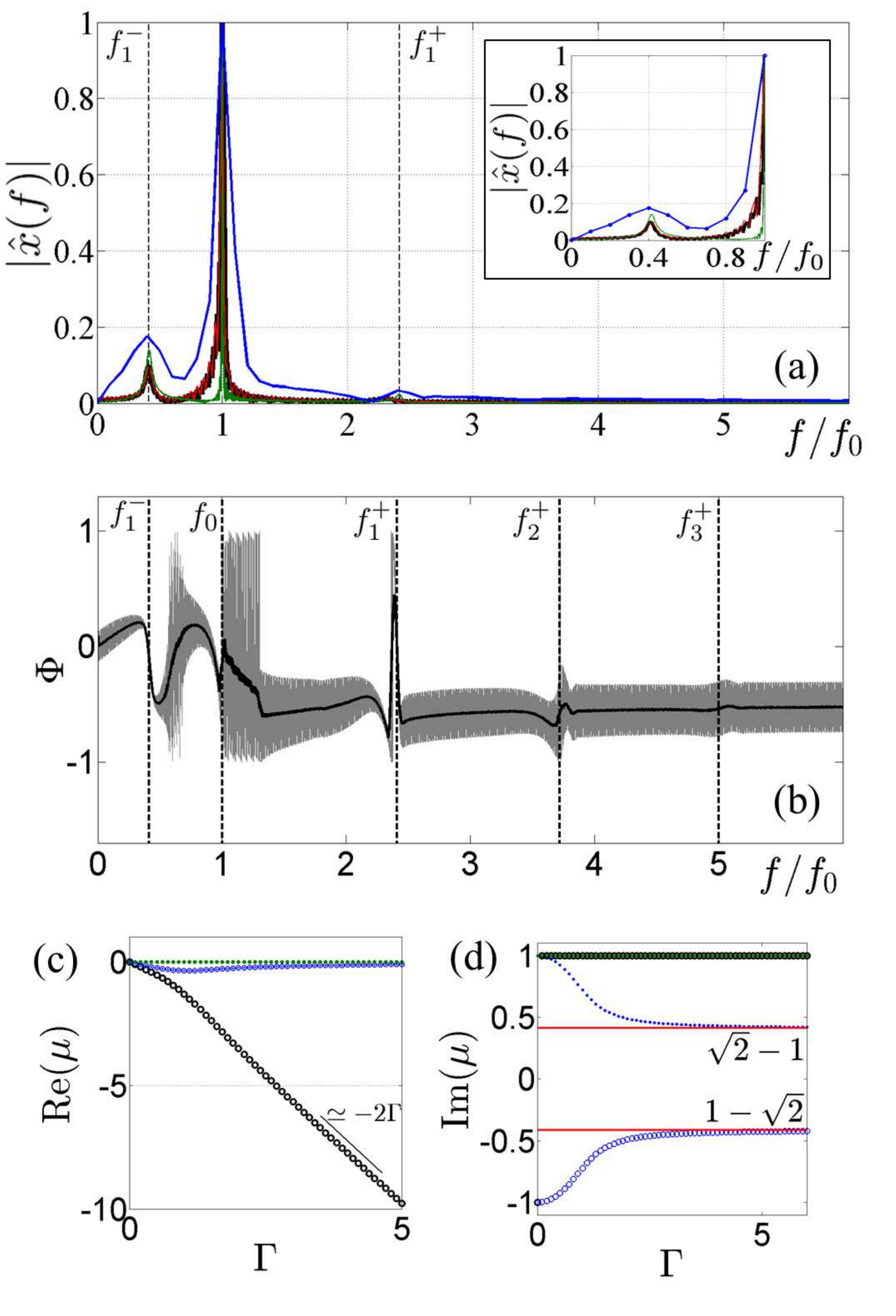} 
\caption{(Color online) (a) Spectrum of the position $x(t)$. The frequencies are normalized by the fundamental frequency $f_0$. Amplitude spectrum $\vert \hat{x}(f) \vert $ of the paths obtained, from experiments [blue (upper) solid line online] from Fort's model (black solid lines) and from the numerical resolution of the Rayleigh oscillator (red solid line online, gray printed, overlapping the black curve) with $\Gamma=25.5$. In green (online, for the printed version downer line at $f/f_0=1$), we plot the spectrum of the transient of Eq. \ref{equationdyninertiaordre1}. In vertical dashed lines are represented the position of the predicted eigenfrequency $f_1^{\pm}=\sqrt{2}\pm1$. (b) The phase spectrum $\Phi$ of $x(t)$ of a transient ruled by the two dimensional Rayleigh equation (in grey), and in the corresponding smoothed curve (in black). We indicate by the vertical dashed line the higher order eigenfrequencies $f_n^{+}=n+ \sqrt{n+1}$ of the Rayleigh oscillator. (c) and (d) Real and imaginary parts of the Floquet coefficients ordered accordingly to their asymptotic value $\Gamma \gg 1$. Black circles $\mu_s \sim -2\Gamma + i$. Green (online, black printed version) points, $\mu_0 = i$. Blue (light gray, printed version) circles and blue (light gray, printed version) points, $\mu=\sim \pm (\sqrt{2}-1)i-O(1/\Gamma)$ giving way to $f^{\pm}_1$ \cite{Ozaprivate}.\label{fig6}}
\end{centering}
\end{figure}

	The energetic approach proves the existence of only two stable limit cycles $v=\pm 1$, but it does not provide any information about the time evolution. The latter can be analyzed in principle by seeking the response to a perturbation in the neighborhood of the fixed points $(x_0,y_0)=(\cos(t+\psi),\pm \sin(t+\psi))$. As the amplitude of the nonlinear term vanishes as $\sim (v-v_0)/v_0$ close to the stable limit cycle, its influence can be probed by a perturbative approach. Introducing a perturbation as $x=\cos t+\epsilon x_1$ and $y=\sin t+\epsilon y_1$ leads to
\begin{equation}
\left\{
    \begin{array}{ll} 
     \displaystyle     \ddot{x}_1+\Gamma\dot{x}_1+x_1  =\Gamma\left(\dot{y}_1\sin 2t + \dot{x}_1\cos 2t \right)\\
     \displaystyle \ddot{y}_1+\Gamma\dot{y}_1+ y_1  =\Gamma\left(\dot{x}_1\sin 2t -\dot{y}_1  \cos 2t  \right).
 
     \end{array}
\right.
\label{equationdyninertiaordre1}
\end{equation}
	The left-hand side corresponds to a damped harmonic oscillator but the meaning of the parameter $\Gamma=1/(\omega T_V)$ differs from a simple damping ratio.  Indeed, Fig. \ref{fig3}b indicates over-damped relaxation with $\Gamma_1=1/(\omega_1 T_V) < \Gamma_m$ while  Fig. \ref{fig3}c shows under-damped oscillatory motions with $\Gamma_2=1/(\omega_2 T_V) > \Gamma_m$. The right-hand side is the first order expansion of the nonlinear coupling of the Rayleigh equation and introduces time-dependent coefficients with a amplitude proportional to the parameter $\Gamma$. This term provides both a periodical forcing on the velocity and a coupling with the velocity along the other directions of space. The solution $\bm Z_1=(x_1,\dot x_1, y_1, \dot y_1 )$ can be represented in a four-dimensional phase space with a time evolution $\dot{\mathbf{Z}}_1=A \mathbf{Z}_1$ prescribed by $A$
\begin{equation}
A(t)=\begin{bmatrix} 
 0 & 1 & 0 & 0 \\ 
 -1 & -\Gamma\left(1 -\cos 2t\right) & 0 & \Gamma\sin 2t \\ 
 0 & 0 & 0 & 1 \\ 
 0 & \Gamma\sin 2t & -1 &  -\Gamma\left(1 +\cos 2t\right)\end{bmatrix}
\end{equation}
The adequate theoretical framework was developed by Floquet \cite{Floquet_1883} and indicates that the solution must be of the form
\begin{equation}
\mathbf{Z}_1=e^{\mu t}\mathbf{g},
\end{equation}
	where $\mathbf{g}$ is a function of time and $\mu$ the Floquet coefficients. The $\pi$ periodicity of the elements of $A$ implies that $\mathbf{g}$ satisfies the condition $\mathbf{g}(t+\pi)=\mathbf{g}(t)$ and that it is also expandable in Fourier series. Here we are only interested in the values of the Floquet coefficients $\mu$ and not the full solution that would be given by the Fourier decomposition of $\mathbf{g}$. A way to find the asymptotic ($\Gamma=25.5 \gg \Gamma_m$) values of $\mu$, corresponding to the experimental situation is to note that the coupling term in the right hand side of Eq. \ref{equationdyninertiaordre1} induces a shift of $\pm 2$ in the frequency of oscillation of $x_1$ and $y_1$. An asymptotic solution can be sought under the form
\begin{equation}
\left\{
    \begin{array}{ll} 
    x_1=-ie^{\mu t}\left(1-ae^{2it} +\ldots\right)\\
    y_1=e^{\mu t}\left(1+ae^{2it}+\ldots\right)
 
     \end{array}
\right.
\end{equation} 
Once inserted in Eq. \ref{equationdyninertiaordre1}, $x_1$ and $y_1$ are solutions only for four particular values of $\mu$. The corresponding frequencies are
\begin{equation}
\left\{
    \begin{array}{ll}
 2\pi f_s=1+2\Gamma i\\ 
 2\pi f_0=1\\
 2\pi f_1^+=1+\sqrt{2}\\
 2\pi f_1^-=1-\sqrt{2}
     \end{array}
\right.
\end{equation}
The Floquet coefficients can be found numerically \cite{Ozaprivate} by first solving  $\dot{U}=AU$, with $U(t=0)$ the $4\times 4$ identity matrix and then finding the eigenvalue of $U(\pi)$, $\lbrace \exp \mu_k \pi \rbrace_{k=1,...,4}$. As observed in Figs \ref{fig6}(c) and \ref{fig6}(d), $f_s$ is found to be always stable as its related Floquet coefficient has a negative real part. The Floquet coefficients of asymptotic values $\mu \sim \pm i(\sqrt{2}-1)$ giving way to the frequency $f_1^{\pm}$, are stable and are asymptotically ($\Gamma \gg \Gamma_m$) neutrally stable [see Fig. \ref{fig6}(c) and \ref{fig6}(d)]. The fourth Floquet coefficient, signature of the rotational invariance of the steady state \cite{Ozaprivate}, is a pure imaginary number, and corresponds to a neutrally stable solution. For this particular case, the linear approach reaches its limit for analyzing the stability of the Rayleigh oscillator. Let us mention that we truncated the initial equation of motion with symmetry arguments; a linear perturbative approach of the complete dynamical equation could recover a linear stability at short memory arising from terms of higher symmetry.    
	The values of these particular frequencies should also appear in any transient to the circular motion since they represent a signature of the nonlinear terms when the dynamics spreads from the limit cycle. This behavior can be observed in the transient regime. Figure \ref{fig6}(a) shows the spectrum density of $x(t)$ during a transient as a function of the normalized frequency $f/f_0$ with $f_0$ the fundamental frequency. They have been computed from an experimental transient, the numerical model, the Rayleigh equation and the linearized system of equations. The four predicted frequencies, $\mathrm{Re}(f_s)$, $f_0$, $f^-_1$, $f^+_1$ can be well identified. For both the numerical model and the Rayleigh equation resolution, others frequencies appear for $f^+_2 = 3.73 \pm 0.02$ and $f^+_3=5 \pm 0.02$. These higher order frequencies are of small amplitude, almost undetectable looking at the amplitude spectrum, but have a well defined signature in the phase spectrum $\Phi$ [See Fig. \ref{fig6}(b)]. In the perturbative development of the Rayleigh equation, i.e., Eq. \ref{equationdyninertiaordre1}, these frequencies do not arise. They correspond to higher order terms, which cannot be revealed by the first order expansion. A development at higher orders of $x=x_0+\epsilon x_1+\epsilon^2 x_2 \ldots$ and $y=y_0+\epsilon y_1+\epsilon^2 y_2\ldots$ would give the whole set of eigenfrequency
\begin{equation}
 f_n^+= n+\sqrt{n+1}
\end{equation}  \\

\section{Conclusion \label{Sec:IV}}
	For the short memory regime, we show that a walker placed in a two-dimensional harmonic potential well converges to a circular motion. The mechanism of convergence involves a dissipation with two features. First, the kinetic energy converges to the equilibrium kinetic energy. Second, the oscillations of the velocity close to its mean value relax with another time scale. The first effect limits the time of convergence for large orbits, whereas the second one is dominant for small orbits. The time corresponding to the transition between these two regimes is typically the time needed to travel a Faraday wavelength. The fluctuations over a distance smaller than the Faraday wavelength will then be guided by this relaxation of speed oscillations. These two transient regimes can be described by a two dimensional Rayleigh oscillator. This proves that the complex underlying hydrodynamic description is here reduced to a standard nonlinear system. From the theoretical point of view, we highlight the stability of the circular attractor by an energetic argument where the linear stability analysis reaches its limit. Only the nonlinear terms are responsible for this stability. Nevertheless the linear expansion predicts the correct set of eigenfrequencies $\lbrace\mathrm{Re}(f_s) ,f_0 ,f^-_1,f^+_1\rbrace$, potentially arising in any perturbation of the walker trajectory. \\
	
	This approach isolates the effect of the propulsion of the walker dynamics from the other contributions, particularly a complex feedback from the memory effects. In the case of a harmonic well, we have thus studied the low memory dynamics near $V=V_0$. For the long memory regime, the trajectories would be much more complex. However, the fluctuations of the norm of the speed are limited so that the symmetry arguments remain relevant in the tangential direction. Consequently the dominant terms of the tangential component of the propulsive force must be of the form of the Eq. \ref{Forcepropulsion}, even when the memory increases. In this sense, the Rayleigh equation is a key ingredient for understanding any fluctuations in velocity as observed in several experiments \cite{Harris_PRE_2013,Perrard_natureC_2013}. 

\acknowledgements{The authors are grateful to Y. Couder and E. Fort for encouraging this work and J. Fronteau, J. Bush, and R. Rosales for useful discussion. The authors are especially grateful to A. Oza for his important remarks about Floquet theory which led to Fig. \ref{fig6}(c) and \ref{fig6}(d). This research was supported by the French Agence Nationale de la Recherche, through the project "ANR Freeflow". This work was supported by LABEX WIFI (Laboratory of Excellence ANR-10-LABX-24) within the French Program "Investments for the Future" under the reference ANR-10-IDEX-0001-02 PSL*.}
\section{Appendix: Propulsion in the short memory regime\label{meca}}

The propulsion force can be alternatively derived from mechanical arguments. In the short memory regime, only the very last rebounds contribute significantly to the surface wave field $h$. For the sake of simplicity, we only retain the contribution of the last bounce. The surface wave field $h$ can be expressed as 
\begin{equation}
h\simeq h_0 J_0\left(k_F\Vert \mathbf{r}(t)- \mathbf{r}(t-T_F) \Vert \right)
\end{equation}  
with $h_0$ the amplitude of the field \cite{Eddi_JFM_2011,Molacek_JFM_2_2013}, $k_F=2\pi / \lambda_F$ the Faraday wave vector and $J_n$ the Bessel function of order $n$. The associated wave force is related to the local slope of the field, $\mathbf{F}_{\mathrm{wave}}=-C\bm{\nabla}h$, with $C$ a coupling constant calculated in \cite{Molacek_JFM_2_2013}. The surface force can be expressed as
\begin{equation}
\mathbf{F}_{\mathrm{wave}}\simeq C h_0 J_1\left(k_FVT_F \right)\dfrac{\mathbf{V}}{\Vert \mathbf{V} \Vert}
\end{equation}  
and be expanded as
\begin{equation}
\mathbf{F}_{\mathrm{wave}}\simeq\dfrac{C h_0}{2} \left( k_FVT_F-\dfrac{(k_FVT_F)^3}{8}\right)\dfrac{\mathbf{V}}{\Vert \mathbf{V} \Vert}
\end{equation}
 and rewritten 
 \begin{equation}
\mathbf{F}_{\mathrm{wave}}\simeq\dfrac{C h_0k_FT_F}{2} \mathbf{V}\left(1-\dfrac{(k_FVT_F)^2}{8}\right)
\end{equation}
The total propulsion force $\mathbf{F}_p$ results from the loss of energy at the surface and the propulsion from the wave and can be expressed as
\begin{equation}
\mathbf{F}_p=-\mu_0 \mathbf{V}+\mathbf{F}_{\mathrm{wave}}
\end{equation}
$\mu_0$ being the apparent friction calculated in \cite{Molacek_JFM_2_2013}. This expression takes the form
\begin{equation}
\mathbf{F}_p=\gamma_0\mathbf{V}\left(1-\dfrac{V^2}{V_0^2} \right)
\end{equation}
with $\gamma_0=k_FT_FCh_0/2-\mu_0$ and $V_0=(16\gamma_0/(C h_0(k_FT_F)^3))^{1/2}$. Beyond the details of the coefficients, the form of this expression is the same as Eq. \ref{Forcepropulsion}. This mathematical expression is relevant if we only consider one last rebound and remains a good approximation in the low memory regime.

\end{document}